# Quantum transport in weakly coupled superlattices at low temperature

E. Lhuillier,[1,2] I. Ribet-Mohamed,[1] A. Nedelcu,[3] V. Berger,[2] E. Rosencher[1]

[1]*ONERA, Chemin de la Hunière, 91761 Palaiseau cedex, France.*

[2]*Matériaux et Phénomènes Quantiques, Université Paris 7, Bat. Condorcet, Case 7021, 75205 Paris cedex 13, France.*

[3]*Alcatel-Thales III-V Lab, Campus de l'Ecole Polytechnique, 1 Avenue A. Fresnel, 91761 Palaiseau cedex, France.*

We report on the study of the electrical current flowing through weakly coupled superlattice (SL) structures under an applied electric field and at very low temperature, i.e. in the tunneling regime. This low temperature transport is characterized by an extremely low tunneling probability between adjacent wells. Experimentally, I(V) curves at low temperature display a striking feature, i.e a plateau or null differential conductance. A theoretical model based on the evaluation of the scattering rates is developed in order to understand this behavior, exploring the different scattering mechanisms in AlGaAs alloys. The dominant interaction in our typical operating conditions is found to be the electron-ionized donors scattering. The existence of the plateau in the I(V) characteristics is physically explained by a competition between the electric field localization of the Wannier-Stark electron states in the weakly coupled quantum wells and the electric field assisted tunneling between adjacent wells. The influence of the doping concentration and profile as well as the presence of impurities inside the barrier are discussed.

PACS number(s): 73.63.Hs, 72.10.-d, 85.60.Gz

## I. INTRODUCTION

Electronic transport in superlattices (SL) has been extensively studied since the early work of Esaki and Tsu[1]. However most studies deal with strongly coupled structures in order to observe high field domain formation[2] or coherence effects such as Bloch oscillations[3]. In this paper we focus on the transport in very weakly coupled SL at low temperature. Only little work has been devoted to the microscopic understanding of this type of tunnel transport[4], despite the observation of new phenomena such as phase transitions[5] and the fact that Quantum Well Infrared Photodetectors (QWIPs) operate in the low coupling regime at low temperature. Most of the existing models are based on the Wentzel-Kramers-Brillouin (WKB) approach. However, as we will show, this model fails to explain the experimental results. We thus developed a microscopic model of transport at low temperature for very weakly coupled SL, based on a scattering approach. This study is of a particularly large scope: indeed, the maturity of GaAs-based materials (low number of defect levels) and the unipolar character of QWIPs (no passivation needed) eliminate unwanted parasitic material effects, and thus only fundamental microscopic interactions are involved in the transport. Our model takes into account six interactions: electron-optical phonon, electron-acoustical phonon, alloy disorder, interface roughness, ionized impurities and carrier-carrier interactions. Due to the very narrow ground miniband we expect that coherent transport[6] and second order effects[7] (two successive tunneling processes via a (virtual) state) stay moderate. As a consequence we investigate hopping transport[8] between ground subbands of adjacent wells[9]. Several papers already addressed this regime[10,11], but generally the coupling between wells investigated by the authors is far larger than ours and their model fails to explain our experimental data. *Our model provides a full quantum description of current transport in weakly coupled SLs, validated by experiments.*

In this paper we first present (sec. II) sample measurements (I(V) curves and spectral response). The I(V) curves at low temperature exhibit in particular a striking null differential conductance (i.e. plateau) behavior. In sec. III, the usual WKB approximation is shown to fail in reproducing this striking feature. Our model, based on the calculation of different scattering rates, is developed in sec. IV. Section V presents the results of our model concerning the scattering rates and the resulting current as a function of the electric field. The dominant interaction in our experimental conditions is found to be the electron-ionized impurity scattering. The existence of the plateau in the I(V) characteristics is explained by a competition between the electric field localization of the Wannier-Stark electron states in the weakly coupled quantum wells and the electric field assisted tunneling. Finnaly, the influence of both the doping profile and the presence of defects in the barrier is presented in Section VI.

## II. EXPERIMENTS

### A. Structure

The experiments have been done on a QWIP structure[12] composed of forty periods with a 73Å wide GaAs well and a 350Å wide $Al_{15.2}Ga_{84.8}As$ barrier. The central third of the well is silicon doped with a concentration of $n_{2D}=3\ 10^{11}cm^{-2}$. The structure is sandwiched between two n-type, silicon doped contacts ($[Si]=10^{18}cm^{-3}$). This QWIP is obtained by MBE growth, and then processed into mesas of 23.5µm lateral size. The barrier is 127meV high and the ground state is located approximately 40meV above the bottom of the GaAs conduction band. The doping value leads to a Fermi level 10.6meV above the ground state.

### B. Measurements

The device was placed on the cold finger of a Janis helium cryostat. The temperature regulation was made with a 330 Lakeshore control unit. Current-voltage measurements were carried out with a 6430 Keithley sub-femtoampere source meter. Special care was dedicated to the fine control of the sample temperature. FIG. 1 (a) presents dark current measurements, which displayed a good repeatability in time and between pixels.

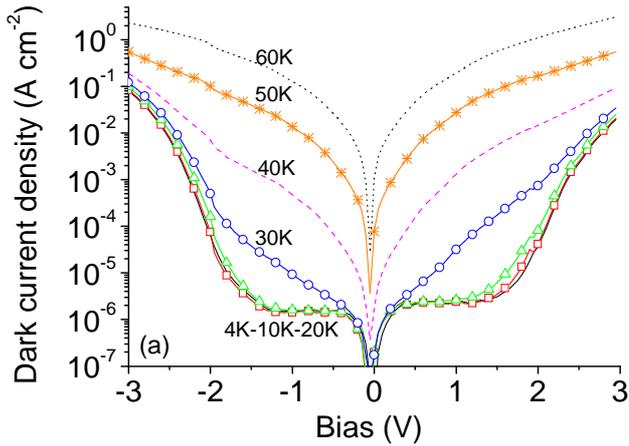

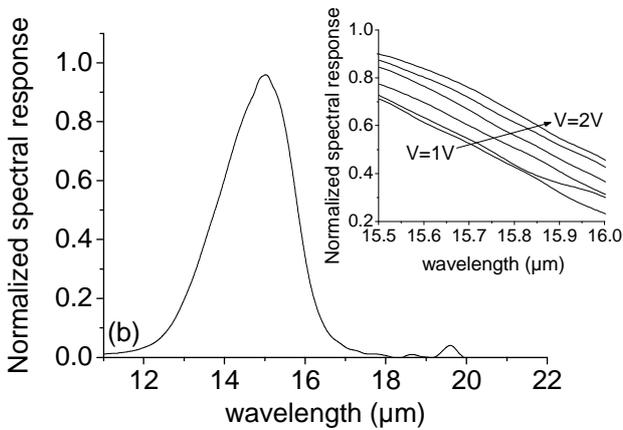

FIG. 1(a) Dark current density as a function of the applied bias for T=4K, 10K, 20K, 30K, 40K, 50K and 60K. (b) Spectral response at T=10K for V=1.5V, inset spectral response for different bias voltages from V=1V to V=2V by step of 0.2V, in the 15.5µm-16µm range.

For T>25K, the current increases monotonously with the temperature. This regime has been extensively studied and is well understood, see for instance ref 13. Below 25K, however, the current is independent on the temperature, which is the sign of the tunneling regime. The low temperature I(V) curve displays three different parts:
- First, an ohmic regime (0V→0.5V) where the current increases linearly with the bias.
- Second, a plateau regime (0.5V→1.5V) where the dependence of the current with the bias is surprisingly low. This plateau is attributed to the transport between the ground states of two adjacent wells. *The main goal of this paper is to explain the very low dependence of the current with the bias in this plateau regime.*
- The high bias regime (V >1.5V) where the current increases very rapidly. This rise is the sign of a change in the transport mechanism. The origin can be attributed to impact ionization[14] in the vicinity of the contact or to transport from the ground state to the continuum in the center of the structure. In the following we will not address this high bias part of the I(V) curve, since this transport mechanism has already been largely investigated in a previous paper[15].

One should notice that the I(V) curves present a slight asymmetry: we will address this effect in section V. Neither hysteresis, nor saw tooth pattern[16] have been observed in our I(V) curves.

The spectral response was measured by a Bruker Equinox 55 Fourier Transfrom InfraRed spectrometer (FTIR) in which the signal is amplified by a Femto – DLPCA 200 amplifier. The measurements are presented in FIG. 1 (b). The QWIP displays a spectral response peaked at 14.5µm, with a full width at half maximum (FWHM) of 2µm. The inset of FIG. 1 (b) shows the variations of the spectral response with the applied bias in the high wavelength part of the spectrum. This point will be further discussed in section III B.

### III. WKB MODELLING

Tunnel transport in QWIP is generally described using the WKB approximation[5,16,17], which relies on two assumptions: (i) The variation of the potential barrier is small compared to the electron wavelength. (ii) The tunneling probability from the final state is negligible.
The WKB expression, which gives the tunneling probability of a particle of energy E through a potential barrier U(x) between points *a* and *b*, is given by the expression (50.9) from the Landau-Lifchitz book[18]:

$$D=\exp\left(\frac{-2}{\hbar}\int_a^b p(x)dx\right)$$

where m* the effective mass of the electron in GaAs, $\hbar$ the reduced Planck constant and

$p(x) = \sqrt{2m^*(E - U(x))}$ is the electron momentum. Such an approximation leads to the following expression for the current density:

$$J_{WKB} = e\frac{m^*}{\pi\hbar^2}\int_{E1}^{\infty}\tau_{WKB}^{-1}(E)f_{FD}(E)dE \quad (1)$$
$$- e\frac{m^*}{\pi\hbar^2}\int_{E1-eFL_b}^{\infty}\tau_{WKB}^{-1}(E)f_{FD}(E)dE$$

where e is the elementary charge, $E_1$ is the ground state energy, $f_{FD}(E) = \left[1 + \exp(\frac{E-E_{fw}}{k_bT})\right]^{-1}$ the Fermi Dirac population factor, $k_b$ the Boltzmann constant, T the temperature, $E_{fw}$ the Fermi level in a well, F the electric field, $L_b$ the barrier width, $\tau_{WKB}^{-1}$ is the inverse of the time for which an electron succeeds in crossing the barrier. Following Gomez[5] $\tau_{WKB}^{-1}$ could be written $\tau_{WKB} = \frac{2L_w}{v}P^{-1}$ in which $L_w$ is the well width, $v$ is the electron speed given by $E = 1/2m^*v^2$ and

$$P = \exp(-\frac{4\sqrt{2m_b^*}}{3.e.F.\hbar} \times [(V_b - E)^{3/2} - (V_b - E - e.F.L_b)^{3/2}]) \quad (2)$$

is the WKB probability that the electron tunnels through the trapezoidal barrier. Here $V_b$ is the barrier height and $m_b^*$ the effective mass in the barrier.

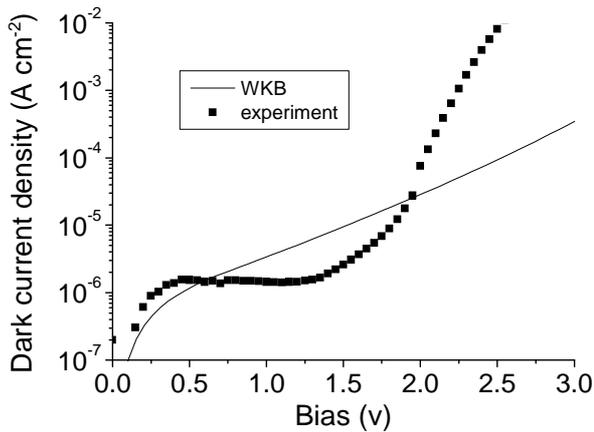

FIG. 2. Experimental and theoretical (WKB) current density as a function of the applied bias.

FIG 2 shows a comparison between WKB prediction and the experimental results. Clearly, WKB approximation fails to reproduce the I(V) plateau. This result is consistent with the probability P in Eq.(2) being a strict monotonic function of the applied bias V.

This discrepancy is however unexpected since WKB approximation generally yields a good agreement with experience for similar devices[5].

The reason of this discrepancy is the following. In Ref 18, it is clearly stated that WKB approximation is valid if one neglects the reflected wave from the final state (x>b) (see the discussion below expression 50.2 in ref 18). This can be easily explained in simple terms. In the WKB approximation, the potential barrier between points *a* and *b* is split into thin slices {$x_i$, $x_{i+1}$} of thickness Δ. *If one neglects the reflected waves in the potential barrier (*i.e. condition ii), the particle wavefunction between $x_i$ and $x_{i+1}$ is: $\psi_i(x) = A_i e^{-\frac{p_i}{\hbar}x}$ so that the probability $D_i$ of the wave to reach $x_{i+1}$ from $x_i$ is $D_i = e^{-2\frac{p_i}{\hbar}\Delta}$ if the variation of $p(x)$ is small over Δ (condition i). The probability of the particle to tunnel through the potential barrier is thus:

$$D = \prod_i D_i = e^{-\frac{2}{\hbar}\sum_i p_i \Delta} = e^{-\frac{2}{\hbar}\int_a^b p(x)dx}$$

As clearly stated in the reference 18, this is valid if the electron wave function is delocalized for x>b, but not if the wavefunction is localized for x<b, in which case the reflected wave from *b* to *a* is of course not negligible. Consequently, the WKB approximation cannot take into account the effect of scattering in the neighbouring wells, which is the main coupling mechanism for transporting the electrons from well to well in this hopping regime.

### IV. SCATTERING APPROACH

We thus chose to develop a scattering approach of transport in multi Quantum Wells (MQW). Scattering methods have already been used to model the quantum transport in heterostructures for resonant tunnel diodes[19], MQW structures[20,21] and more recently in Quantum Cascade Lasers[22,23] (QCLs). But only little work has been devoted to applying this method to weakly coupled SLs[24], mainly because of the difficulty to deal with the low coupling effects.

The high quality of the GaAs material, grown by Molecular Beam Epitaxy (MBE), allows us to evaluate a scattering rate and a current from microscopic Hamiltonians, since no uncontrolled or detrimental material defects (deep levels, hopping on defects,…) prevail. Our model includes the six main interactions observed in GaAs-based materials: optical phonon, acoustical phonon, alloy disorder, interface roughness, ionized impurities and interactions between carriers (see appendix A to F for details on the scattering rate evaluation). No *a priori* hypothesis is made concerning the magnitude of each process. However, we assume that the GaAs material grown by MBE is of high enough quality to disregard scattering due to dislocations. We also assume that no neutral impurities are involved in the transport mechanism.

It is important to understand that the tunnel transport between ground states is a very inefficient mechanism in the weakly coupled Quantum wells (QW) considered here. Indeed, we can assume that a MQW is a stack of doped planes with a typical doping of $3 \cdot 10^{11} cm^{-2}$. Considering that the current density in the plateau regime is $10^{-6} A.cm^{-2}$, (see FIG. 1a) we can conclude that the typical scattering rate is given by:

$$\tau = \frac{e.n}{J} \approx \frac{1,6.10^{-19} \; 3.10^{11}}{10^{-6}} \approx \text{some ten ms} \quad (3)$$

This means that an electron is scattered from one well to the next every ten milliseconds. This time should be compared to the intra-well scattering time which is less than one picosecond[20,25] within the conduction band, (ten orders of magnitude smaller). *Consequently we are dealing with very unlikely events.*

Our model is based on the evaluation of the inter-well scattering times $\Gamma^{-1}$ using the Fermi Golden Rule (FGR). The tunnel transport between ground states is rather simple to model since it only couples two dimensional (2D) levels. The time $\Gamma^{-1}$ is included in the current expression:

$$J = \int_{E_1}^{\infty} \frac{em^*}{\pi\hbar^2} . \Gamma(E,F).(1-f_{FD}(\varepsilon_f))f_{FD}(\varepsilon_i)d\varepsilon_i, \quad (4)$$

where $\frac{m^*}{\pi\hbar^2}$ is the 2D density of states (DOS). The use of an equilibrium population factor is motivated by the fact that the inter-well scattering rate is several orders of magnitude lower than the intra-well rate, which leads to a thermalized subband for each well[26]. Our model includes the direct current (J+, the electron relaxes from the upper well to the lower one) and the reverse current[4] (J-, the electron flows up the structure), see FIG. 3. In this Wannier-stark approach, the current writes:

$$J_{Wannier-Stark} = J^+ - J^- =$$
$$= \int_{E_1}^{\infty} e \frac{m^*}{\pi\hbar^2} . \Gamma(E,F).(1-f_{FD}(\varepsilon_f)).f_{FD}(\varepsilon_i)d\varepsilon_i$$
$$- \int_{E_1-eFL_d}^{\infty} e \frac{m^*}{\pi\hbar^2} . \Gamma(E,F).(1-f_{FD}(\varepsilon_f)).f_{FD}(\varepsilon_i)d\varepsilon_i \quad (5)$$

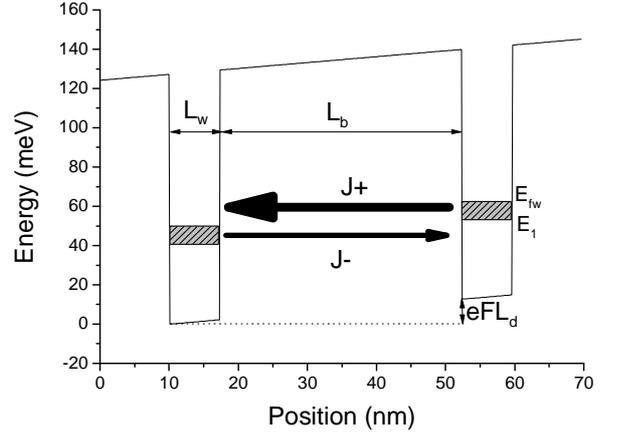

FIG. 3. Band profile of the QWIP under an electric field of 3 kV.cm$^{-1}$.

The expression of the scattering rate given by the FGR is:

$$\Gamma(K_i) = \frac{2\pi}{\hbar} \sum_{Kf} \left|\langle f|\hat{H}|i\rangle\right|^2 \delta(\varepsilon_i - \varepsilon_f) \quad (6)$$

In this expression i and f point out the initial and final states, $\varepsilon_{i/f}$ the associated energy and $\hat{H}$ the perturbation Hamiltonian. The wave functions are evaluated in the envelope function formalism[27] $|i\rangle = |K_i, k_{iz}\rangle \propto \xi_{k_iz}.e^{i\vec{k}\vec{R}}$, z being the direction of the growth. The wavefunctions $\xi_{k_{i/f}z}$ are evaluated using a two band kp method in a two wells structure[28]. The energy associated with this level is $\varepsilon_i = E_i + \frac{\hbar^2 K_i^2}{2m^*}$ with $E_i$ the energy of the ground state. Non-parabolicity for the in-plane dispersion is neglected in the current calculation. Indeed the exchanged energies remain very low compared with the inverse of the coefficient of non-parabolicity of GaAs ($E(1+\alpha E) = \frac{\hbar^2 k^2}{2m^*}, \alpha = 0.61 eV^{-1}$)[29]. The wave functions and the ground state energies are evaluated for each value of the electric field, so that Stark effects are taken into account in our model. The electric field is denominated by F and the period of the superlattice is $L_d$. The periodicity allows us to replace the $E_i - E_f$ quantity by $eFL_d$.

In order to compare the theoretical J(F) curve with the experimental J(V), we assume that the electric field on the structure is homogeneous. It is well known that the electric field distribution leads, for a given bias, to a higher electric field in the vicinity of the contact than in the center of the structure[30,31]. Typically the difference between the homogeneous electric field and the "real" electric field is about a few ten %[32]. Nevertheless, the higher the number of periods, the lower the associated correction. Our structure contains forty periods and this "mean field" approach should be adequate. There are two main consequences to this homogeneous electric field hypothesis: first we neglect all contact effects[33] and then we assume that no electric field domain[13,34] exists in the QWIP. To justify the last point, we used the high wavelength part of the FTIR measurement (Fig. 1 (b)), (assuming that the electric field profile is the same with and without photon flux). In the high wavelength part of the spectrum, the photon energy is lower than the bound-to-extended state transition energy, so that the electron does not have enough energy to be excited directly into the continuum. In fact, the electron is rather subject to tunneling assisted by photon and electric field, through the triangular part of the barrier[35]. This tunneling probability depends on the electric field value. Thus, the translation of the photocurrent spectrum with the bias reflects the field reigning on each quantum well, which allows us to conclude that the bias is effectively applied on the QWIP.

Because of the large barrier involved in our structure, quantum wells are very weakly coupled and the miniband width is in the nano-eV range[15], whereas the potential drop per period is some tens of meV. Electrons are thus highly localized and their wavefunctions are consistently described by their unperturbed quantum well wavefunctions. Consequently our approach is based on a hopping mechanism from one well to the next one. This is an important difference with the paper of Castellano[15] et al in which the I(V) plateau is attributed to a saturation of the electronic velocity in a very narrow miniband (Esaki-Tsu approach).

## V. RESULTS
### A. Parameters used for modeling

A temperature of 10 K is used. The other parameters used for the evaluation of the scattering rates are given in the following table:

TABLE 1. Interaction parameters used for simulation (see Appendices A to F)

| Parameter | unit | value |
|---|---|---|
| $m^{*}$ [20] | kg | $0.067 \cdot m_0$ |
| $\varepsilon_r \approx \varepsilon_s$ [13,48] | $A^2 s^4 m^{-3} kg^{-1}$ | $12.9\,\varepsilon_0$ |
| $\varepsilon_\infty$ [48] | $A^2 s^4 m^{-3} kg^{-1}$ | $10.9\,\varepsilon_0$ |
| $\hbar w_{LO}$ [48] | meV | 36.6 |
| $\rho$ [19] | Kg.m$^{-3}$ | 5320 |
| $c_s$ [19] | ms$^{-1}$ | 5220 |
| $D_c$ [19] | eV | 12 |
| $\Delta$ [21,53,36,37] | nm | 0.3 |
| $\xi$ [21,53,36,37] | nm | 6.5 |
| $\Delta V = V_{AlAs} - V_{GaAs}$ [21] | eV | 0.836 |
| $a$ [48] | nm | 0.565 |
| $V_b = \Delta V \cdot x$ [13] | eV | 0.128 |

with $m_0$ the free electron mass and $\varepsilon_0$ the vacuum permittivity.

### B. Scattering rates

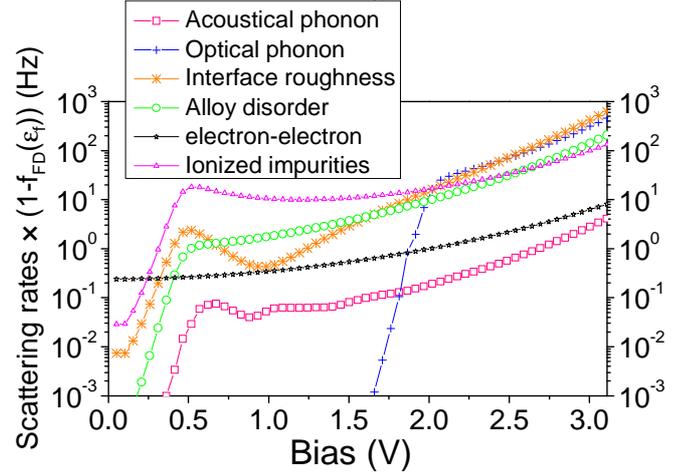

FIG. 4. Product of the scattering rates, for $K_i=0$, by the population factor of the arrival level, as a function of the bias for the six considered processes.

FIG. 4 shows the product $\Gamma(E,F) \times (1 - f_{FD}(\varepsilon_f))$ of the scattering rates, for a null initial wave vector ($K_i=0$), by the population factor of the arrival level, as a function of the bias. One of the main results of this graph is the fact that at low field (V<2V) the dominant interaction is the one between the electrons and the ionized donors. In the plateau regime (0.5V<V<1.5V), this interaction is at least one order of magnitude higher than the others. At higher bias (V>2V), other interactions such as LO phonon and alloy disorder also become important. Concerning LO phonon, we have to underline that this effect will not happen in a higher wavelength QWIP because of smaller exchanged energy.

One should also notice that for very low bias (V<0.5V) the product of the scattering rate by the population factor is increasing, which is the consequence of the $(1 - f_{FD}(\varepsilon_f))$ factor. Electron-electron interactions do not show this behavior because this factor was not included, due to the difficulty to evaluate the energy of the arrival states[23].

The FIG. 4 shows that a plateau is theoretically obtained in the scattering rates vs bias, which will lead to a constant current in this range. One might wonder what the physical origin of this plateau is. In fact, the plateau regime results from the competition between two different effects of the electric field. On the one hand, as described above, an increase of the electric field tends to enhance the wave function in the neighbouring well, see FIG. 5 (c), enhancing the scattering and thus the electrical current. On the other hand, the increase of the electric field tends to localize the wave function in each well (Wannier-Stark effect[2]), see FIG. 5 (b), leading to a decrease of the electrical current, i.e. a negative differential resistance (see FIG. 5).

### C. Level broadening

When the electric field is very low in the structure, the electron wavefunctions, which are unperturbed in our first order perturbation theory, tend to be degenerate and delocalized. This leads to an unrealistic infinite conductivity as it is well known in transport theory[38]. Rott et al.[39] and Wacker[40] have already demonstrated that for Wannier-Stark hopping a $1/F^n$ law is expected at low field, where the n value depends on the considered Hamiltonian. To correctly describe the ohmic regime it is of course necessary to take into account the decoherence effects on the transport mechanism. The question of the decoherence may be treated using a non equilibrium Green's function method, but this method is highly computationally demanding[41]. Other teams have also tried to include decoherence using the density matrix formalism, see the work of Iotti et al.[42], Callebaut et al.[43] and more recently Gordon et al.[44]. In order to take this effect into account while keeping a simple first order calculation, the easiest way is to introduce a lifetime broadening[38].

Following a Wannier-Stark approach[2], the delocalized part of the wave functions magnitude is given by $J_1\left(\frac{\Delta_0}{2eFL_d}\right)$, with $J_1$ the first order Bessel function (see FIG. 5 (b)). The current density is proportional to the part of carrier wave function delocalized in the next well. So the associated current is $J_{\text{Wannier-Stark}} \propto J_1\left(\frac{\Delta_0}{2eFL_d}\right)^2$. In the Wannier-Stark approach the role of the electric field is to localize the wave function when the field is higher than $\Delta_0$. When $F \to 0$, this latter expression diverges as explained above. To take into account the level broadening due to intrawell scattering, we introduce a imaginary part $i\frac{\hbar}{\tau}$ to the transition energy[45] which leads to an effective field $eF_{\text{eff}} L_d = eFL_d + i\frac{\hbar}{\tau}$. It may be easily shown that it is also equivalent to introducing a coherence length for the electron wave function. The value of the dephasing time $\tau$ has been taken equal to the intrawell scattering time (scattering between two states of the same subband and the same well) $\frac{1}{\tau} = 1.1 \times 10^{13}$ Hz and has been obtained with the same scattering method. Such a value is consistent with previous theoretical[20] and experimental[25] results. This value is also very close to the broadening energy (50fs) extracted from our spectral measurements. The expression of the current is thus:

$$J \propto J_1\left(\frac{\Delta_0/2}{\left|eFL_d + i\frac{\hbar}{\tau}\right|}\right)^2 \quad (7)$$

Using the fact that $eFL_d \gg \Delta_0$, and $J_1(x) \underset{0}{\propto} \frac{x}{2}$, one finds:

$$J = J_{\text{Wannier-Stark}} \frac{(eFd)^2}{(eFd)^2 + \left(\frac{\hbar}{\tau}\right)^2}.$$

From this latter expression, it is clear that the diverging effect of the Wannier-Stark delocalisation is smoothed out by the dephasing time $\tau$.

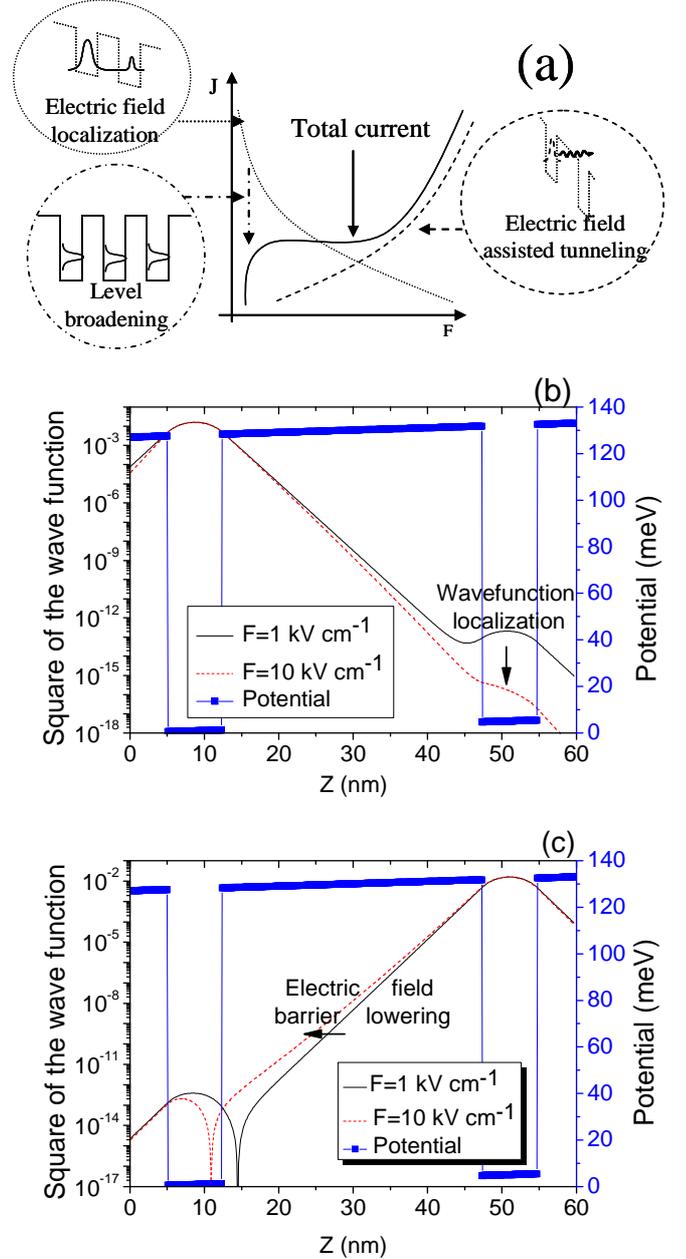

FIG. 5 (a) The current variation for an increasing electric field is the result of a competition between the enhanced probability of the electron to be in the neighbouring well (enhancing scattering), and an enhanced Wannier-Stark localisation of the electrons in their well. For very low fields, the dephasing time

(equivalently the coherence length) of the electrons localize the electron in the wells. (b) Effect of the electric field on the downstream wave function, the arrow shows the effect of the localization on the wave function. (c) Effect of the electric field on the upstream wave function, the arrow shows the effect of the barrier lowering on the wave function.

D. Theoretical dark current

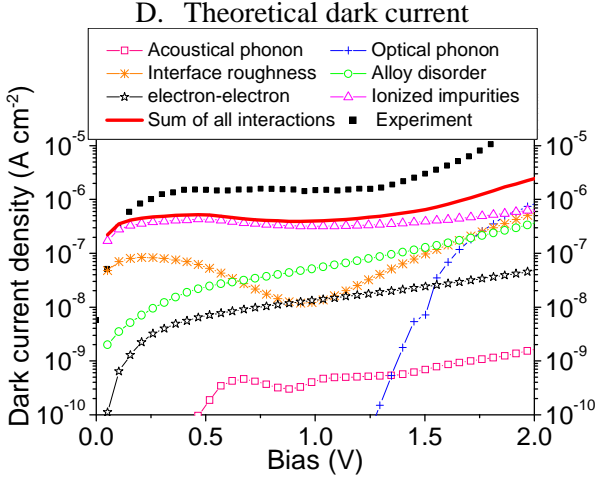

FIG. 6. Current density as a function of the bias for the six processes considered.

FIG. 6 shows the predicted current as a function of the electric field for the six interactions. The theoretical curve taking into account the 6 mechanisms (full line) can be compared with the experimental curve (squares). Our model is able to reproduce the null differential conductance behavior. We can observe that on the plateau regime the agreement between theory and measurement is quite good, typically a factor three. Because of the quadratic dependence with the doping of the current due to ionized donors (see Appendix E) this factor three may result of an uncertainty of "only" 70% on the doping value. Other effects such as uncertainty on aluminum concentration or segregation of aluminum and silicon may also be involved in this difference.

## VI. INFLUENCE OF THE DOPING DENSITY AND PROFILE

One of the main advantages of our microscopic approach is that we can describe the effects linked to the doping density and profile. Such effects are expected to be quite important since we have demonstrated that the electron-ionized donors interaction dominates the plateau regime. We underline that our model gives a quadratic dependence of the current with the doping density:

$$J \propto e\, n\, \Gamma(n) \qquad (8)$$

with an explicit sum over the number of electrons (equal to the doping) and one implicit sum included in the scattering rate over all the scattering centers.

To experimentally validate this dependence of the current as a function of the doping magnitude we have grown two series of samples which only differ by the magnitude of the doping. The first structure is very close to the previous sample with a well (barrier) width of 7.2 nm (34 nm), the aluminium content in the barrier is 15% and a Si doping in the central part of the well. The sheet densities are repectively $1\times10^{11}\mathrm{cm}^{-2}$ (component $B_1$) and $2\times10^{11}\mathrm{cm}^{-2}$ (component $B_2$). The structure includes sixty periods. FIG. 7 presents the associated dark current. The ratio of the two plateau magnitudes is 4.6, whereas 4 was expected. Similar results have been obtained in a second structure ($L_w$=8nm, $L_b$=40nm, %Al=13%, forty periods and a doping of $2\times10^{11}\mathrm{cm}^{-2}$ and $4\times10^{11}\mathrm{cm}^{-2}$), where a ratio very close to 4 was effectively measured.

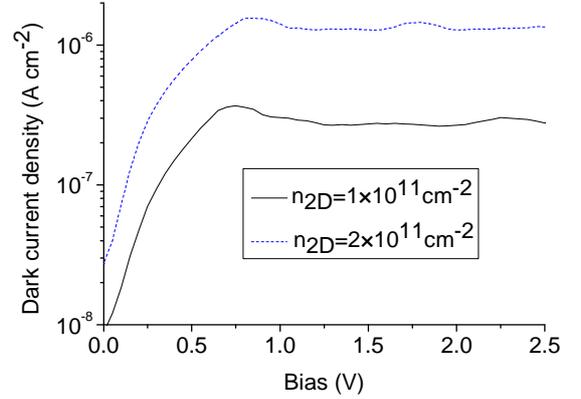

FIG. 7 Dark current as a function of the applied bias for component $B_1$ (doping level of $1\times10^{11}\mathrm{cm}^{-2}$) and $B_2$ (doping level of $2\times10^{11}\mathrm{cm}^{-2}$).

There is a clear added value relatively to the Esaki Tsu-like[15] model which is independent of the doping. Even more, the current is not only sensitive to the doping value, but also to its position.

This last part will study effects such as segregation or the influence of impurities inside the barrier, since those two parameters are difficult to control precisely and may have significant effect on the current.

### A. Doping position

We have first theoretically studied the dependence of the current as a function of the position of these impurities. We have scanned the position of an ideal delta doping (1Å) trough the well and plot the associated current in FIG. 8. The electron sheet density is kept constant through the scan. In our model, we expect the current to decrease while driving the doping layer away from the wavefunction maximum, since the scattering overlap integrals are strongly reduced. This behavior is clearly observed in FIG. 8. Let us note that, because of the applied electric field of 10kV.cm$^{-1}$, the maximum of the curve is not at the center of the well, but shifted of nearly 15 Å in the direction of the electric field.

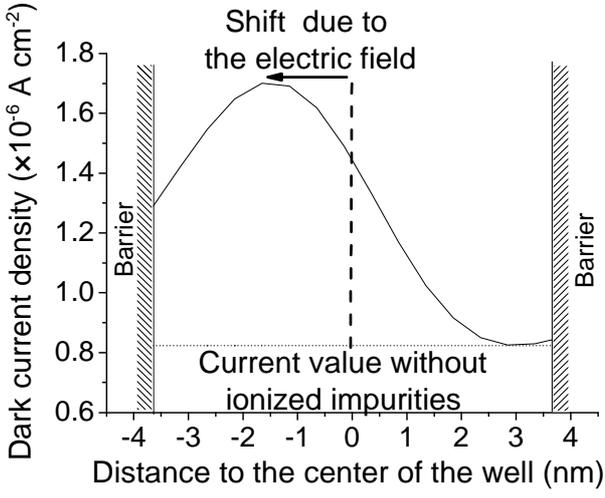

FIG. 8. Dark current density (full line) as a function of the doping position in the well, under an electric field of 10 kVcm$^{-1}$. The dotted line is the current value for the same electric field but without ionized impurities.

Thus it may be possible to reduce the dark current by shifting the doping position from the center of the well to an other position. To confirm this prediction, we have grown two samples ($L_w$=6.8nm, $L_b$=39nm, %Al=15%, 40 periods, sheet density $3\times10^{11}$cm$^{-2}$). The doping is respectively in the central third of the well (component $C_1$) and on the last third (grating side) of the well ($C_2$).

We have plotted on FIG. 9, the magnitude of the dark current as a function of the temperature under a bias of -1.5V. Using X ray diffraction and spectral measurements we have measured that the structure $C_2$ presents a lower confinement (transition energies respectively of 88.3 meV for C1 and 87.4 meV for C2). Consequently at high temperature (T> 35 K), this sample presents a higher thermoionic dark current due to a more efficient thermal activation of the electron. At low temperature however, in spite of its lower confinement, this C2 structure displays a less important tunnel current, which results from the fact that the interwell hopping scattering rate has been effectively reduced, as expected fro our theory.

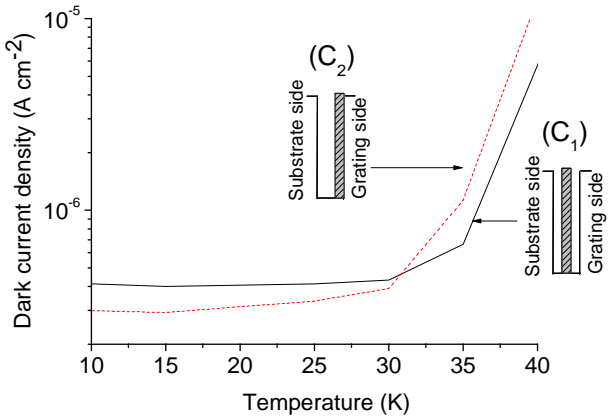

FIG. 9 Dark current as a function of the temperature, under a voltage bias of -1.5V, for the component $C_1$ and $C_2$. The grey pattern indicates the doping position into the well.

### B. Segregation and I(V) asymmetry

Since our model takes into account the doping profile, it can also be used to predict the effect of doping segregation[46]. The segregation length is a function of the growth temperature[47] and partial pressure of the different deposited elements. As a realistic approximation[31], we assumed that our doping distribution is an asymmetric trapezoid. This trapezoid is composed of three zones, see the inset of FIG. 10.

The first zone of length $L_1$ represents the segregation in the direction opposite to the growth. This segregation is quite low and thus the segregation length ($L_1$) is taken equal to 5 Å.

The second zone corresponds to the nominal place of doping.

The last zone corresponds to a segregation in the direction of the growth, and consequently shows a higher segregation length ($L_2$). For our typical growth temperature the segregation length is in the range 25-50Å, as reported by Wasilewski et al[47].

The volume doping density has been chosen such that, whatever the values of the segregation lengths, the sheet density remains unchanged.

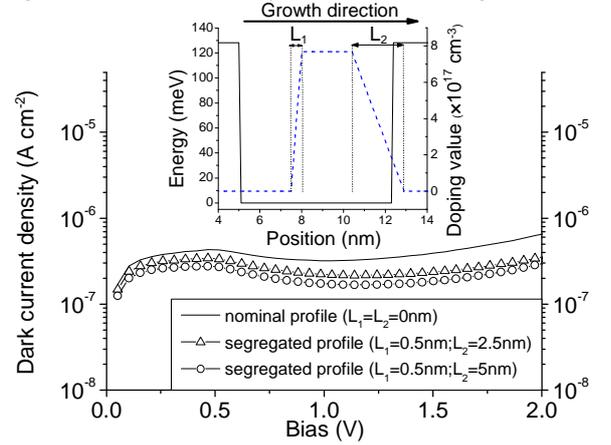

FIG. 10. Theoretical dark current density, as a function of the electric field, for different values of segregation lengths ($L_1$ and $L_2$). Inset: doping profile in a quantum well.

We observe that the segregation reduces the dark current (FIG. 10). This is explained by the fact that the segregation tends to move the doping away from the center of the well.

The asymmetry of the I(V) curves is often attributed to the doping segregation. Rather than changing the polarity of the applied field in our simulation (which implies to re-evaluate all wave functions and energy values), we changed the direction of segregation, $L_1$ becoming $L_2$ and vice versa.

Using our segregated profile ($L_1$=5Å, $L_2$=50Å), our model predicts a ratio of I$^+$ (current under positive bias) over I$^-$ (current under negative bias of 1.3, for a bias of 1V (on the plateau). The experimental value of this ratio is included in the 1.5-1.6 range, leading to a difference between experimental and theoretical value of 20%.

As expected, the segregation introduces an asymmetry with the bias polarity. The quantitative agreement is acceptable if we consider the hypothesis made on the doping profile shape.

### C. Importance of the growth method.

As we have shown that the ionized impurities play a major role in the value of the dark current, it is important to study the influence of the growth method. Indeed, because of the high reactivity of the aluminum[54], non desired impurities can be present in the barrier. Impurities such as carbon, oxygen, silicon, sulfur, tellurium and germanium can be incorporated with a concentration which is dependent on the growth method. With MBE the residual concentration is below SIMS resolution,[54] typically ~some $10^{14} cm^{-3}$. With Metal Organic Chemical Vapor Phase Epitaxy (MOVPE) this concentration is typically one order of magnitude higher. As shown in FIG. 11, the presence of these impurities in the barrier, added to the nominal doping, has no influence for MBE, and is also negligible in the MOVPE case. Such a result is very important for the QWIP designer, since both methods can be used without major impact on the device performances.

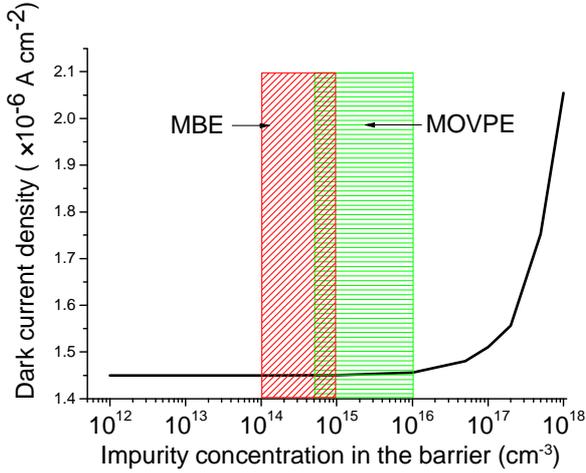

FIG. 11. Dark current density as a function of the concentration of undesired ionized impurities in the barrier under an electric field of 10kVcm$^{-1}$. The two rectangles highlight the typical range of concentration for MBE (leaned pattern) and for MOVPE (horizontal pattern)

### VII. CONCLUSION

We have studied the electronic transport under dark condition of weakly coupled QWs at very low temperature, i.e. in the tunneling regime. The I(V) curves exhibit a plateau region, where the current displays a very low dependence with respect to the applied electric field. We have checked that this does not originate from electric field domain effects. We have shown that the usual WKB approximation is unable to reproduce this striking plateau regime. Consequently we developed a full quantum scattering approach of the transport, based on the Fermi golden rule and taking into account all the main interactions met in AlGaAs heterostructures. Our model suggests that the plateau regime is due to a competition between two mechanisms when the electric field is enhanced in the QWs: a decrease of the current due to the electric field localization of the carriers (Wannier-Stark effect) and an increase due to a higher scattering probability due to an increasing tunnel effect. We conclude that the electron-ionized donors interaction is the dominant one and obtain a good agreement between theory and experiment for the plateau value. We have applied our model to predict the influence of the doping density and profile on the dark current. We showed that our model is able to reproduce the I(V) curves asymmetry, at low bias, by the use of a segregated doping profile. We also demonstrated the very low effect of the choice of the growth method on the dark current. This work may promote the development of new doping profiles for QWIP operating in the low photon flux regime.

### AKNOLEDGEMENTS

Authors thank Angela Vasanelli and Robson Ferreira for fruitful discussions.

### APPENDIX A: ELECTRON-LONGITUDINAL OPTICAL (LO) PHONON

The typical value of the energy drop per period (10 to 35 meV in the plateau regime), remains lower than the GaAs LO phonon energy (36meV). Consequently there is a very low probability that this mechanism is the main one, at least for F<10kV.cm$^{-1}$. Moreover the very low operating temperature is not favorable to this type of scattering.

The Hamiltonian describing the interaction between electron and longitudinal phonon can be written as [19,20] $H_{e^-/phonon} = \sum_q \alpha(q) e^{i\vec{q}\vec{r}} b_q^+ + hc$, where q is the momentum of the phonon, $\alpha(q)$ describes the strength of the interaction and $b_q^+$ the phonon creation operator. In the particular case of optical phonon $\alpha(q)$ is linked to the Frölich interaction[48], where the electric field due to the dipole of the GaAs doublet interacts with an electron:

$$\alpha_{LO}(q)^2 = \frac{e^2 \hbar w_{LO}}{2\Omega \varepsilon_p} \frac{1}{q^2 + q_0^2}, \quad (A1)$$

with $\frac{1}{\varepsilon_p} = \frac{1}{\varepsilon_\infty} - \frac{1}{\varepsilon_s}$. Here $\hbar w_{LO}$ is the energy of the optical phonon is GaAs, $\Omega$ the volume of the sample, $\varepsilon_\infty$ and $\varepsilon_s$ are respectively the dielectric constant at infinite and null frequency and q$_0$ the inverse of the screening length. Such an Hamiltonian considers only bulk phonons. Readers interested in other types of phonons (surface phonon for example) could read ref 49 and 50. The matrix element is equal to:

$$M_{LO}{}^2 = \alpha_{LO}{}^2 \sum_q \frac{1}{q^2 + q_0{}^2} \left| \langle k_f | e^{\pm iqr} | k_i \rangle \right|^2$$

$$= \sum_{q_z, Q} \frac{\alpha_{LO}{}^2}{q^2 + q_0{}^2} \left| \langle \xi_f | e^{iq_z z} | \xi_i \rangle \right|^2 \left| \langle e^{iK_f r''} | e^{iQr''} | e^{iK_i r''} \rangle \right|^2$$

(A1)

Here we can define the form factor related to this interaction $F_{LO}(q_z) = \left| \langle \xi_f | e^{iq_z z} | \xi_i \rangle \right|^2$. Now this expression can be included in the FGR giving:

$$\Gamma_{LO}(K_i) = \frac{\alpha_{LO}{}^2}{4\pi^2 \hbar} \iiint K_f dK_f d\theta dq_z \frac{F_{LO}(q_z)}{q^2 + q_0{}^2}$$

$$\times \delta(E_i - E_f + \frac{\hbar^2}{2m^*}(K_i{}^2 - K_f{}^2) - \hbar w_{LO})$$

(A2)

By evaluating the integral over $K_f$ one finds[23]:

$$\Gamma_{LO}(K_i) = \frac{\alpha_{LO}{}^2 m^*}{(2\pi)^2 \hbar^3} \int d\theta \int dq_z \frac{F_{LO}(q_z)}{q^2 + q_0{}^2}$$ (A3)

where

$$q^2 = Q^2 + q_z{}^2,$$ (A4)

and

$$Q^2 = K_i{}^2 + K_f{}^2 - 2K_i K_f \cos(\theta)$$ (A5)

and

$$K_f{}^2 = K_i{}^2 + \frac{2m^*}{\hbar^2}(E_i - E_f - \hbar w_{LO}).$$ (A6)

In such an expression, the Bose Einstein factor which describes the population of phonons is taken equal to unity. This choice is justified by the fact that temperature is very low, which implies that no optical phonon absorption is possible.

## APPENDIX B: ELECTRON-LONGITUDINAL ACOUSTICAL (AC) PHONON

As for optical phonons, the very low temperature is again not favorable to the interaction between acoustical phonons and electrons. However, because of their lower energy, our model needs to include both emission and absorption of acoustical phonons.

The considered Hamiltonian is a classical electron-bulk acoustical phonon interaction, based on Debye dispersion i.e. $H_{e^-/phonon} = \sum_q \alpha(q) e^{i\vec{q}\vec{r}} b_q^+ + hc$ [19]

with:

$$\alpha_{AC}(q)^2 = \frac{\hbar D_c{}^2}{2\Omega \rho . c_s} q$$ (B1)

where $D_c$ is the acoustic deformation potential, $\Omega$ the volume of the sample, $\rho$ the density and $c_s$ the sound velocity, q the phonon wave vector and $b_q^+$ the phonon creation operator. We define $\alpha_{AC}{}^2 = \frac{\hbar D_c{}^2}{2\rho . c_s}$ [19] for an easier reading. The matrix element associated with this interaction can be written as:

$$M_{AC}{}^2 = \sum_q \alpha_{AC}{}^2 q \left| \langle f | e^{iqr} | i \rangle \right|^2$$ (B2)

or

$$M_{AC}{}^2 = \sum_{q_z, Q} \alpha_{AC}{}^2 q \left| \langle \xi_f | e^{iq_z z} | \xi_i \rangle \right|^2 \left| \langle e^{iK_f r''} | e^{iQr''} | e^{iK_i r''} \rangle \right|^2$$

(B3)

The form factor associated to this interaction is defined by the following expression $F_{AC} = \left| \langle \xi_f | e^{iq_z z} | \xi_i \rangle \right|^2$. Finally the transition rate is given by [23]

$$\Gamma_{AC} = \frac{\alpha_{AC}{}^2}{4\pi^2 \hbar^2 c_s} \int_0^\infty K_f dK_f \int_0^{2\pi} d\theta \frac{q^2}{q_z}(1 + n_{BE}(\hbar w)).F_{AC}(q_z)$$

(B4)

with

$$q = \frac{E_i - E_f + \frac{\hbar^2}{2m^*}(K_i{}^2 - K_f{}^2)}{\hbar c_s},$$ (B5)

and

$$Q^2 = K_i{}^2 + K_f{}^2 - 2K_i K_f \cos(\theta)$$ (B6)

and

$$q_z = \sqrt{q^2 - Q^2}$$ (B7)

Here $n_{BE}$ is the Bose Einstein distribution. In the case of absorption, this $1+n_{BE}$ factor is replaced by $n_{BE}$.

## APPENDIX C: ALLOY DISORDER (AL)

In $Al_x Ga_{1-x} As$ alloy the presence of aluminium in substitution of the gallium induces scattering, because of the different atomic potential of the two atoms. It is quite hard to evaluate *a priori* the magnitude of this interaction. Such a scattering is usually treated by a potential proportional to the deformation. $V = \Delta V . \delta x(r)$ [21] where $\Delta V$ is the band offset between GaAs and AlAs. Generally, the calculation consists in defining a statistical correlation function between the aluminium atoms positions, following the Nordheim rule[51]:

$$\langle \delta x(r) \delta x(r') \rangle = \Omega_0 [x(1-x)] \delta(r - r')$$ (C1)

where $\Omega_0$ is the size of the primitive cell. We can now evaluate the mean value of the matrix element

$$\langle M_{AL} \rangle^2 = \left\langle \left| \langle i | \Delta V \delta x | f \rangle \right|^2 \right\rangle$$ (C2)

and so

$$\langle M_{AL}\rangle^2 = \Delta V^2 \left\langle \left| \Psi_f^*(r)\delta x(r)\Psi_i(r)d^3r \right|^2 \right\rangle \quad (C3)$$

We define the associated form factor

$$F_{AL} = \int_{alloy} |\xi_f(z)|^2 |\xi_i(z)|^2 dz. \quad (C4)$$

To conclude, the expression of the scattering rate is given by:

$$\Gamma_{AL} = \frac{\Delta V^2 \Omega_0 [x(1-x)] F_{AL}}{2\pi\hbar} \int \frac{m^*}{\hbar^2} d\varepsilon_f \int d\theta \delta(\varepsilon_i - \varepsilon_f), \quad (C5)$$

which is equal to

$$\Gamma_{AL} = \frac{m^*}{\hbar^3} \Delta V^2 \Omega_0 [x(1-x)] F_{AL}. \quad (C6)$$

### APPENDIX D: INTERFACE ROUGHNESS (IR)

Because of the wide barriers and the small number of interfaces we do not expect this process to be dominant. This is a main difference between QWIP and Quantum Cascade Detector[52] (QCD) or QCL. In the latter, wells are highly coupled and barriers are quite thin, making interface roughness a non negligible interaction at low temperature[22]. The treatment for interface roughness[21,53] is very close from the one made for alloy disorder: Unuma[21] makes the remark that interface roughness is the sheet equivalent of the alloy disorder. As for alloy disorder we start by defining a linear potential with the perturbation:

$$V_{IR}(r) = V_b \Delta \delta(z - z_i) F(r), \quad (D1)$$

where $V_b$ is the band offset between GaAs and $Al_xGa_{1-x}As$, $\Delta$ is the magnitude of the interface defects and $F(r)$ is the spatial distribution of defects. The delta function underlines the local character of this interaction. Most often $F(r)$ is chosen to follow a Gaussian correlation function:

$$\langle F(r).F(r')\rangle = \exp\left(-\frac{|r-r'|^2}{\xi^2}\right) \quad (D2)$$

with $\xi$ the correlation length. The matrix element is now given by

$$\langle M_{IR}\rangle^2 = \left\langle \left| \langle f | V_b.\Delta.\delta(z-z_i)F(r) | i \rangle \right|^2 \right\rangle \quad (D3)$$

or

$$\langle M_{IR}\rangle^2 = V_b^2 \Delta^2 \left\langle \left| \int d^3r \Psi_f^*(r)\Psi_i(r)\delta(z-z_i)F(r) \right|^2 \right\rangle \quad (D4)$$

and the very simple form factor can be written:

$$F_{IR} = |\xi_i(z_i)|^2 |\xi_f(z_i)|^2. \quad (D5)$$

In the case of multiple interface the form factor is summed over all interface positions ($z_i$). Considering the elastic character of the interaction

$$K_j^2 = K_i^2 + \frac{2m^*}{\hbar^2}(E_i - E_f), \quad (D6)$$

the exchanged wave vector $Q = K_i - K_f$ becomes:

$$Q^2 = 2K_i^2 + 2\frac{m^*}{\hbar^2}(E_i - E_f) - 2K_i\sqrt{K_i^2 + 2\frac{m^*}{\hbar^2}(E_i - E_f)}\cos(\theta). \quad (D7)$$

To finish, the scattering rate is:

$$\Gamma_{IR} = \frac{m^* V_b^2 \Delta^2 \xi^2 F_{IR}}{2\hbar^3} \int_0^{2\pi} e^{-\frac{Q^2\xi^2}{4}} d\theta. \quad (D8)$$

### APPENDIX E: IONIZED IMPURITIES (II)

Scattering by ionized impurities is involved in two different ways in QWIP: first, through the doping which is generally localized in the well. Then the high reactivity of the aluminum in the barrier leads to the inclusion of undesirable impurities (mostly carbon) in the barrier. Concentration of residual impurities is highly dependent of the growth method[54]. As shown in Sec. VI-C, our model can conclude on the importance of residual impurities. It is also able to take into account the segregation of the doping, which leads to asymmetric I(V) curves with bias polarity.

In the following we assume that impurities are completely ionized, even at low temperature. The Coulombian Hamiltonian is:

$$V(r) = \frac{e^2}{4\pi\varepsilon_0\varepsilon_r}\frac{1}{r}. \quad (E1)$$

V is Fourier transformed[39]

$$V(Q) = \frac{e^2}{2\varepsilon_0\varepsilon_r}\sum_Q \frac{e^{-Q|z-z_i|}e^{iQ(r''-r_i'')}}{Q}. \quad (E2)$$

Thus the matrix element is given by

$$M_{II}^2 = \left(\frac{e^2}{2\varepsilon_0\varepsilon_r}\right) \left| \int dz d^2r'' \times \int d^2Q \xi_f^*(z) e^{-iK_f r''} \frac{e^{-Q|z-zi|}e^{iQ(r''-ri'')}}{Q} \xi_i(z)e^{iK_i r''} \right|^2 \quad (E3)$$

We can also define the associated form factor

$$F_{II}(Q) = \left| \int dz \xi_f^*(z) e^{-Q|z-zi|} \xi_i(z) \right|^2, \quad (E4)$$

so we get $M_{II}$ by

$$M_{II}^2 = \frac{e^4}{4\varepsilon_0^2 \varepsilon_r^2} \frac{F_{II}(K_i - K_f)}{|K_i - K_f|^2}. \quad (E5)$$

The scattering rate is obtained by summing over all positions of the doping ($z_{ii}$):

$$\Gamma_{II} = \frac{e^4}{8\pi\hbar\varepsilon_0^2\varepsilon_r^2} \frac{m^*}{\hbar^2} \int_{impurities} dz_{ii} N(z_{ii})$$
$$\times \int d\theta \frac{F_{II}(\sqrt{|K_i - K_f|^2 + q_0^2})}{|K_i - K_f|^2 + q_0^2} \quad (E6)$$

where

$$|K_i - K_f| = \sqrt{K_i^2 + K_f^2 - 2K_iK_f \cos(\theta)} \quad (E7)$$

and

$$K_f^2 = K_i^2 + \frac{2m^*}{\hbar^2}(E_i - E_f). \quad (E8)$$

In order to take into account the screening of the interaction, we used a Thomas-Fermi approach with a constant screening length[48] $q_0^2 = \frac{e^2 n}{\varepsilon_r k_b T}$, n the volumic doping and $\varepsilon_r$ the permittivity of the materials. It leads to an effective wave vector[39,55] $q_{eff} = \sqrt{q^2 + q_0^2}$. Note that unlike the previous process (LO, AC, AL, IR), Coulombian interactions will lead to a quadratic dependence with the doping.

## APPENDIX F: ELECTRON ELECTRON (EE)

Electron-electron interaction is certainly the most difficult interaction to understand. To deal with it, some authors use Green function formalism[56,57]. Only little work has been devoted to the treatment of this interaction using the envelop formalism, we could quote works of Smet[58], Harrison[23] and Kinsler[59]. This lack is the consequence of the very time consuming numerical treatment. All theoretical difficulties in the treatment of this interaction are due to the two bodies type of this interaction. Initial states will be noted as $|i\rangle$ and $|j\rangle$, and the final states as $|f\rangle$ and $|g\rangle$. The interaction potential is, as for ionized impurities, the Coulombian potential, so the matrix element can be written

$$M_{ee} = \left\langle \frac{\xi_f(z)e^{ik_f r''}}{\sqrt{A}} \frac{\xi_g(z')e^{ik_g r'''}}{\sqrt{A}} \right| \frac{e^2}{4\pi\varepsilon_0\varepsilon_r} \frac{1}{r}$$
$$\left| \frac{\xi_i(z)e^{ikr''}}{\sqrt{A}} \frac{\xi_j(z')e^{ikr'''}}{\sqrt{A}} \right\rangle \quad (F1)$$

After having Fourier transformed the potential and defined the form factor by:

$$F^{ee}_{ijfg}(q_{xy}) = \iint \xi_f^*(z)\xi_g^*(z')\xi_i(z)\xi_j(z')$$
$$\times e^{-q_{xy}|z-z'|} dz dz', \quad (F2)$$

the matrix element $M_{ee}$ becomes

$$M_{ee} = \frac{e^2}{2\varepsilon_0\varepsilon_r A q_{xy}} F^{ee}_{ijfg}(q_{xy}) \delta(k_f + k_g - k_i - k_j). \quad (G3)$$

This expression could be injected in the FGR to obtain the expression of the scattering rate:

$$\Gamma^{ee} = \frac{2\pi}{\hbar} \sum_{f,g} \left| \frac{2\pi e^2}{4\pi\varepsilon_0\varepsilon_r A} \frac{F^{ee}_{ijfg}(q_{xy})}{q_{xy}} \right|^2$$
$$\times \delta(k_f + k_g - k_i - k_j)\delta(\varepsilon_f + \varepsilon_g - \varepsilon_i - \varepsilon_j) \quad (F4)$$

This expression depends on the kinetic energy of the two initial states $|i\rangle$ and $|j\rangle$. In order to use the scattering rate of the process in the same way as the previous ones, we sum this expression over all j initial states.

$$\Gamma^{ee} = \frac{e^4}{2\pi\hbar(4\pi\varepsilon_0\varepsilon_r)^2} \iiint \left|\frac{F^{ee}_{ijfg}(q_{xy})}{q_{xy}}\right|^2 P_{j,f,g}(k_j,k_f,k)$$
$$\times \delta(k_f + k_g - k_i - k_j)\delta(\varepsilon_f + \varepsilon_g - \varepsilon_i - \varepsilon_j) dk_j dk_f dk_g \quad (F5)$$

where

$$P_{j,f,g} = f_{FD}(\varepsilon_j)(1 - f_{FD}(\varepsilon_f))(1 - f_{FD}(\varepsilon_g)) \quad (F6)$$

is the population factor of the different states which appear in the expression. Because it is very difficult to obtain separately $\varepsilon_f$ and $\varepsilon_g$, the expression is simplified into $P_{j,f,g}(k_j,k_f,k) = f_{FD}(\varepsilon_j)$. This approximation allows us to obtain an upper limit of the scattering rate, which is not an issue if this interaction is not the main one. Kinsler et al[59] have however proposed a solution to avoid this approximation. To finish the expression of the scattering rate is given by

$$\Gamma^{ee} = \frac{m^* e^4}{(4\pi\hbar)^3 \varepsilon_0^2 \varepsilon_r^2} \int_0^{2\pi}\int_0^{2\pi} \left|\frac{F^{ee}_{ijfg}(q_{xy})}{q_{xy}}\right|^2$$
$$\times P_{j,f,g}(k_j,k_f,k_g) k_j dk_j d\alpha d\theta \quad (F7)$$

with $\alpha$ and $\theta$ two angles. The expression of $q_{xy}$ is the following one

$$q_{xy}^2 = \frac{2k_{ij}^2 + \Delta k_o^2 - 2k_{ij}\sqrt{k_{ij}^2 + \Delta k_o^2}\cos\theta}{4} \quad \text{(F8)}$$

with

$$k_{ij}^2 = k_i^2 + k_j^2 - 2k_i k_j \cos\alpha \quad \text{(F9)}$$

and

$$\Delta k_o^2 = \frac{4m^*}{\hbar^2}(E_i + E_j - E_f - E_g) \quad \text{(F10)}$$

For a more detailed calculation one should read Harrison's book[23].

[1] L. Esaki and R. Tsu, IBM J. Res. Develop. **14**, 61 (1970).
[2] A.Wacker, in *Theory of transport properties of semiconductor nanostructures* (Chapman & Hall, London, 1998), chap. 10.
[3] K. Leo, P. Haring Bolivar, F. Brüggemann, R. Schwedler and K. Köhler, Solid States Communications **84**, 943 (1992).
[4] H. Willenberg, O. Wolst, R. Elpelt, W. Geißelbrecht, S. Malzer and G.H. Döhler, Phys. Rev. B **65**, 035328 (2002).
[5] A. Gomez, V. Berger, N. Péré-Laperne and L.D. Vaulchier, App. Phys. Lett. **92**, 202110 (2008).
[6] R.F.Kazarinov and R.A. Suris, Sov. Phys. Semiconductors **6**, 120 (1972).
[7] H.Willenberg, G.H. Döhler, J. Faist, Phys. Rev. B **67**, 085315 (2003).
[8] M.Pollack, B.Shklovskii, in *Hopping transport in solids* (North Holland, Amsterdam, 1991).
[9] R. Tsu and Dohler, Phys. Rev. B **12**, 680 (1975).
[10] D. Calecki, J. Palmier and A. Chomette, Journal of Physics C: Solid State Physics **17**, 5017 (1984).
[11] S. Rott, N. Linder and G. Dohler, Superlattices and Microstructures **21**, 569 (1997).
[12] E. Lhuillier, I. Ribet-Mohamed, M. Tauvy, A. Nedelcu, V. Berger, E. Rosencher, Infrared Phys. Tech. **52**, 132 (2009).
[13] H. Schneider and H.C. Liu, in *Quantum well infrared photodetectors – Physics and applications* (Springer, Heidelberg, 2006).
[14] L. Gendron, V. Berger, B. Vinter, E.Costard, M. Carras, A. Nedelcu and P. Bois, Semiconductor Science and Technology **19**, 219 (2004).
[15] F. Castellano, F. Rossi, J. Faist, E. Lhuillier and V. Berger, Phys. Rev. B **79**, 205304 (2009)
[16] B.F. Levine, J. Appl. Phys. **74**, R1 (1993).
[17] L. Thibaudeau, P. Bois and J.Y. Duboz, J. Appl. Phys. **79**, 446 (1996).
[18] L. Landau and E. lifchitz in *Mécanique quantique, théorie non relativiste* (Edition Mir, Moscow, 2nd edition 1967).
[19] F. Chevoir and B. Vinter, Phys. Rev. B **47**, 7260 (1993).
[20] R. Ferreira and G. Bastard, Phys. Rev. B **40**, 1074(1989).
[21] T. Unuma, M. Yoshita, T. Noda, H. Sakaki and H. Akiyama, J. Appl. Phys. **93**, 1586 (2003).
[22] A. Leuliet, A. Vasanelli, A. Wade, G. Fedorov, D. Smirnov, G. Bastard and C. Sirtori, Phys. Rev. B **73**, 085311 (2006).
[23] P. Harrison, in *Quantum wells, wires and dots: Theoretical and Computational Physics of Semiconductor Nanostructures*, first edition (Wiley interscience, Chichester UK, 2000).
[24] N. E. I. Etteh and P. Harrison, J. Quan. Elec. **37**, 672 (2001).
[25] A. Alexandrou, V. Berger, and D. Hulin, Phys. Rev. B **52**, 4654 - 4657 (1995).
[26] Luis L. Bonilla and H.T. Grahn, Rep. Prog. Phys **68**, 577 (2005).
[27] G. Bastard, in *Wave Mechanics Applied to Semiconductor Heterostructures* (Les Editions de Physique, Paris, 1988).
[28] C. Sirtori, F. Capasso, J. Faist, and S. Scandolo, Phys. Rev. B **50**, 8663 (1994).
[29] M. Braun and U. Rossler, J. Phys. C **18**, 3365 (1985).
[30] E. Rosencher, F. Luc, P. Bois and S. Delaitre, Appl. Phys. Lett. **61**, 468 (1992).
[31] M. Ershov, V. Rizhii, C. Hamaguchi, Appl. Phys. Lett. **67**, 3147 (1995)
[32] Eric Costard, Ph.D. thesis, University Paris XI, France, 1991.
[33] E. Rosencher, F. Luc, P. Bois and S. Delaitre, Appl. Phys. Lett. **61**, 468 (1992).
[34] H. Schneider, C. Schönbein, R. Rehm, M. Walther and P. Koidl, Appl. Phys. Lett. **88**, 051114 (2006).
[35] J.L. Rouzo, I. Ribet-Mohamed, N. Guérineau, R. Haïdar, M. Tauvy, E. Rosencher, and S.L Chuang, Appl. Phys. Lett **88**, 091117 (2006).
[36] H. Sakaki, T. Noda, K. Hirakawa, M. Tanaka and T. Matsusue, Appl. Phys. Lett. **51**, 23 (1987).
[37] R. Gottinger, A. Gold, G. Abstreiter, G. Weimann and W. Schlapp, Euro. Lett. **6**, 183 (1988).
[38] K. Seeger in *Semiconductor Physics, an introduction* (Springer, Heidelberg, 9th edition, 2004)
[39] S. Rott, N. Linder, G.H. Döhler, Phys. Rev. B **65**, 195301 (2002).

[40] A. Wacker, Physics Reports **357**, 1 (2002).
[41] S.C. Lee and A. Wacker, Phys. Rev. B 66, 245314 (2002).
[42] R. C. Iotti, E. Ciancio and F. Rossi, Phys. Rev. B **72**, 125347 (2005).
[43] H. Callebaut and Q. Hu J. Appl. Phys 98, 140505 (2005).
[44] A. Gordon and D. Majer, Phys. Rev. B **80**, 195317 (2009).
[45] C. Cohen-Tannoudji, B. Diu and F. Laloë, in *Quantum mechanics*, (Hermann, Paris, 1997).
[46] M. Carras, V. Berger, X. Marcadet and B. Vinter, Phys. Rev. B **70**, 233310 (2004).
[47] Z. Wasilewski, H.C. Liu and M. Buchanan, J. Vac. Sci. Tech. B **12**, 1273 (1994).
[48] E. Rosencher and B. Vinter, in *Optoélectronique*, $2^{nd}$ edition (Dunod, Paris, 2002).
[49] Multilayer/DCM electron-phonon scattering : Fermi golden rule calculation
P. Kinsler, http://www.kinsler.org/physics, on 12/11/2008.
[50] H. Rücker, E. Molinari, P. Lugli, Phys. Rev. B **45**, 6747 - 6756 (1992).
[51] C. Kittel, in *Solid state physics*, $7^{th}$ edition, (Wiley, Chichester UK, 1996).
[52] L. Gendron, C. Koeniguer, V. Berger, and X. Marcadet, Appl. Phys. Lett. **86**, 121116 (2005).
[53] T. Unuma, T. Takahashi, T. Noda, M. Yoshita, H. Sakaki, M. Baba and H. Akiyama, Appl. Phys. Lett. **78**, 3448 (2001).
[54] C. Asplund, H. Malm and H. Martijn, Proc. SPIE **6206**, 62060F (2006).
[55] P.J. Price, Surf. Science **113**, 199 (1982).
[56] K. Kempa, Y. Zhou, J. R. Engelbrecht, P. Bakshi, H. I. Ha, J. Moser, M. J. Naughton, J. Ulrich, G. Strasser, E. Gornik and K. Unterrainer, Phys. Rev. Lett. **88**, 226803 (2002).
[57] K. Kempa, Y. Zhou, J.R. Engelbrecht and P. Bakshi, Phys. Rev. B **68**, 085302 (2003).
[58] J. H. Smet, C. G. Fonstad and Q. Hu, J. Appl. Phys. **79**, 9305 (1996).
[59] P. Kinsler, P. Harrison and R.W. Kelsall, Phys. Rev. B **58**, 4771 (1998).